# Determination of Surface Morphology of TiO$_2$ nanostructure using synchrotron X-ray standing wave technique

Gangadhar Das[1,2,*], Manoj Kumar[3], A. K. Biswas[1], Ajay Khooha[1], Puspen Mondal[1] and M. K. Tiwari[1,2]

[1]*Indus Synchrotrons Utilization Division, Raja Ramanna Centre for Advanced Technology, Indore-452013*
[2]*Homi Bhabha National Institute, Anushaktinagar, Mumbai-400094, India*
[3]*Laser Materials Processing Division, RRCAT, Indore-452013, India*
*E-mail: rnrrsgangadhar@gmail.com*

**Abstract.** Nanostructures of Titanium oxide (TiO$_2$) are being studied for many promising applications, *e.g.*, solar photovoltaics, solar water splitting for H$_2$ fuel generation etc., due to their excellent photo-catalytic properties. We have synthesized low-dimensional TiO$_2$ nanoparticles by gas phase CW CO$_2$ laser pyrolysis. The laser synthesis process has been optimized for the deposition of highly pure, nearly mono-dispersed TiO$_2$ nanoparticles on silicon substrates. Hard x-ray standing wave-field (XSW) measurements in total reflection geometry were carried out on the BL-16 beamline of Indus-2 synchrotron radiation facility in combination with x-ray reflectivity and grazing incidence x-ray fluorescence measurements for the determination of surface morphology of the deposited TiO$_2$ nanostructures. The average particle size of TiO$_2$ nanostructure estimated using transmission electron microscopy (TEM) was found to closely agree with the XSW and grazing incidence x-ray diffraction (GIXRD) results.

## INTRODUCTION

Recently, a considerable extent of research is being actively devoted for the development of advanced functional materials for efficient utilization of solar energy, as a long term sustainable energy resource [1-2]. Titanium oxide (TiO$_2$) nanoparticles (NP) are considered as one of the most promising photo-catalyst candidate for these applications [3]. Lack of adequate knowledge about the growth kinetics of nucleation and formation of a layer structure makes it difficult to tune the overall process from the point of view of practical device applications of these low dimensional, highly pure, crystalline TiO$_2$ nanomaterials. The conventional surface microscopy techniques are usually unable to provide any depth dependent information about the physical and chemical states of the nanoparticles. Therefore, one requires suitable methodologies that can be used for to synthesizing different nanostructures effectively on the large surface area of a substrate with a high reproduction rates and reliable means to characterize such complex nanostructures. X-ray standing wave (XSW) induced grazing incidence x-ray fluorescence (GIXRF) combined with x-ray reflectivity (XRR) measurements provide sensitive information about the nature of NP dispersion, agglomeration, average vertical size, shape as well as internal structure and chemical composition of nanoparticle over a large surface area of the substrate [4].

Here, we report synthesis of TiO$_2$ nanoparticles using CW CO$_2$ laser pyrolysis process [5] for the photocatalytic applications and their non-destructive fast surface characterization on a substrate surface. The surface morphology of the TiO$_2$ nanostructure has been investigated using combined XRR and GIXRF measurements. Transmission electron microscopy (TEM) and grazing incidence x-ray diffraction (GIXRD) measurements were also carried out for the deposited TiO$_2$ nanostructures to evaluate surface coverage and crystalline structure of the particles. The results obtained from TEM measurements confirm the findings of XSW analysis on the TiO$_2$ nanoparticle.

## EXPERIMENTAL

TiO$_2$ nanostructures were synthesized using an in-house developed CW CO$_2$ laser based gas phase pyrolysis setup. We used C$_2$H$_4$ as sensitizer gas and Ti(OC$_3$H$_7$)$_4$ as a precursor material for the fabrication of NP. The TiO$_2$ NP were fabricated under the inert environment with argon, as a cooling gas medium at constant pressure of ∼ 650 mbar in the reaction chamber on top of a Si (100) substrate. Different experimental parameters of the deposition process have been varied to optimize the experimental conditions for the formation of spherical shape and well dispersed nanoparticles. The simultaneous XRR and grazing incidence x-ray fluorescence (GIXRF) measurements were carried out on BL-16 beamline of Indus-2 synchrotron radiation facility [6] to characterize the surface morphology of TiO$_2$ nanostructure using 8keV incident x-rays. A commercial TEM (Philips, CM200) with W-filament as cathode was used for the estimation of average particle size and surface coverage of TiO$_2$ nanoparticles. We have also performed GIXRD measurements using a commercial BRUKER D-8 x-ray reflectometer station at Cu-K$\alpha$ wavelength to investigate the crystalline structure of the TiO$_2$ nanostructure.

## RESULTS AND DISCUSSIONS

XSW induced GIXRF intensities are extremely sensitive to the nature of surface structure of a material *i.e*. if it is a thin film, nano, macro particle or buried layer etc. Fig. 1(a) depicts simulated [4] GIXRF profiles of the TiO$_2$ nano-spheres comprising of different various particle sizes, whereas in Fig 2(b) we have plotted simulated GIXRF profiles for TiO$_2$ thin film of various thicknesses. All the simulations were carried out at 8keV incident x-ray energy. In Fig 2(b) we have also plotted Si-K$\alpha$ fluorescence intensity emitted from the Si substrate at 8 keV incident x-ray energy. It is evident from these figures that fluorescence yields (FY) emitted in the case of a nanoparticle and a thin film are markedly different from one another. From these figures one can clearly identify a transition region near $\theta \approx 0.25°$ where the characteristic feature of the nanoparticles layer is seemingly transformed into a thin continuous layer. Fig. 2(a) and (b) represents the measured and fitted combined XRR and GIXRF profiles of TiO$_2$ nanostructures at 8 keV incident x-ray energy respectively. It is evident from Fig. 2(b) that the measured fluorescence profile of Ti-K$\alpha$ matches quite well with the simulated profile when a three layer model comprising of a SiO$_2$ layer on top of Si substrate, a bare TiO$_2$ thin layer, and a low density TiO$_2$ NP layer on top of the TiO$_2$ thin film is assumed for fitting experimental Ti-K$\alpha$ yield. For the comparison purpose we have also plotted computed GIXRF profiles for the individual also (assume only nanoparticle layer and thin film). It can be seen that these profiles significantly differs from the measured Ti-K$\alpha$ GIXRF profiles. However, the XRR profile could be fitted considering a very low dense SiO$_2$ overlayer of thickness 2.4nm. The XRR profile shows the critical angle of Si and it does not show any signature for the presence of TiO$_2$ NP and TiO$_2$ layer. This can be attributed to the fact that the TiO$_2$ layer is quite thin (d ~ 4.5 nm) having a very low density (0.4 gm/cm$^3$). The incident x-ray beam is unable to distinguish the low electron density contrast between low density TiO$_2$ layer and Si substrate. GIXRF profile shows a strong peaking behavior in the vicinity of the critical angle of Si substrate ($\theta_c \approx 0.22°$) due to the coincidence of the XSW antinodes (see Fig. 2d) at the location of NP. It may be noted here that the XRR technique gives information only for the reflecting layers (SiO$_2$ and TiO$_2$ thin layers). On the other hand, GIXRF technique gives information for the non-reflecting layer of nanoparticle as well as for the reflecting TiO$_2$ layer. Thus, the combined (XRR & GIXRF) analysis gives us a novel capability to completely resolve the structure of a complex layered medium comprising of both particles as well as reflecting layer structure. In our case the best fit results indicate the presence of TiO$_2$ nano-spheres of average diameter $\approx 8.0 \pm 1.0$ nm with an average density of $\rho \approx 0.04$ gm/cm$^3$ with a underneath, a TiO$_2$ thin film layer of thickness d ~ 4.5 nm, density $\rho \approx 0.4$ g/cm$^3$ on top of the Si substrate. The roughness values for NP layers and TiO$_2$ thin film were found to be $\sigma$~0.6 nm and $\sigma$~0.8nm respectively.

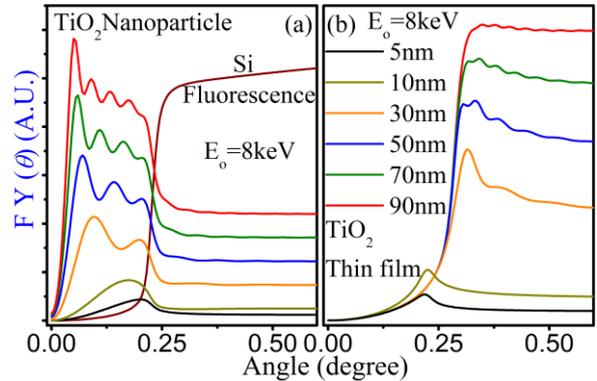

**FIGURE 1**. Calculated XSW assisted GIXRF profiles at 8 keV incident x-ray energy for two cases: (a) TiO$_2$ nano-spheres of various particle sizes, and (b) TiO$_2$ thin film of different film thicknesses on top of Si substrate.

The average particle size of the TiO$_2$ nanospheres was also evaluated using the grazing incidence x-ray diffraction measurements using Scherrer's formula and was found to be in the range of ~ 8-10 nm (see the inset of Fig. 2b) corresponding to (101) reflection. The GIXRD measurements shows anatase phase of the NP with tetragonal crystal structure. Figure 2(c) shows the TEM micrograph of the TiO$_2$ NP. The estimated particle`s size was found in the range of 7 nm to10 nm. In figure 2(d), we have shown the normalized x-ray field intensity distribution on the Si substrate at the 8 keV incident x-ray energy.

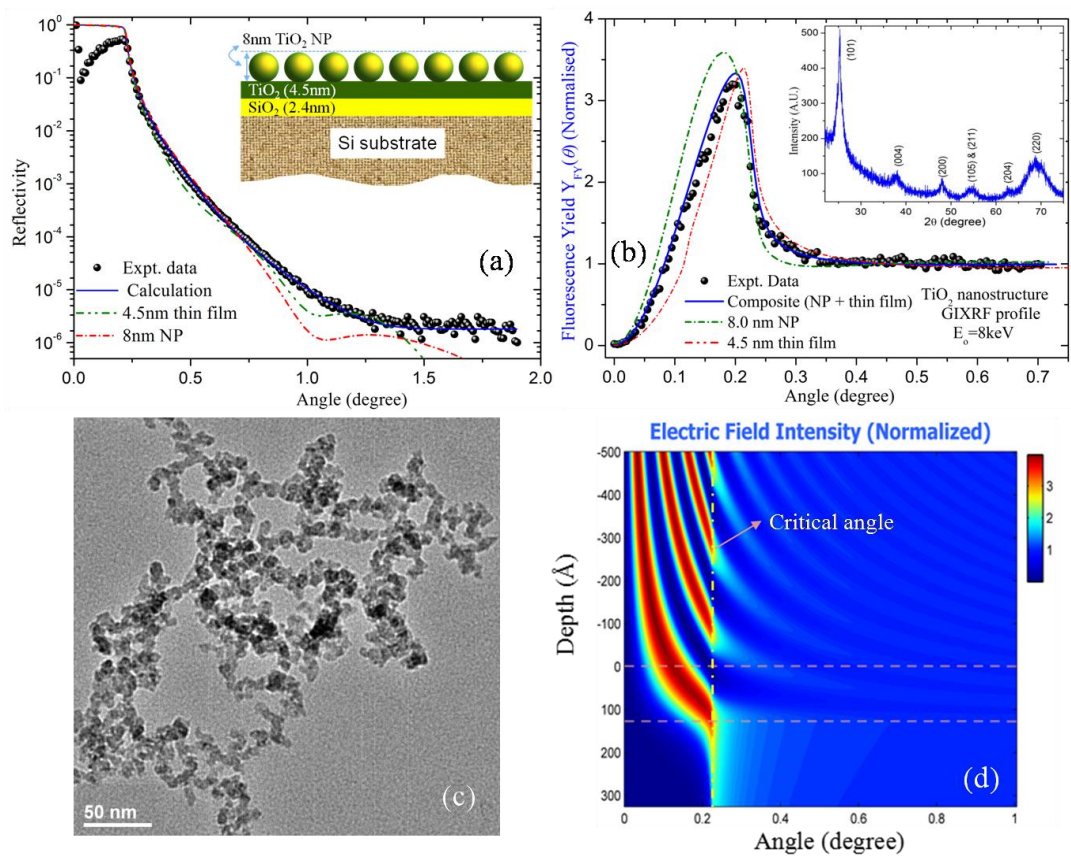

**FIGURE 2.** Measured and simulated (a) XRR and (b) XSW induced GIXRF profiles of TiO$_2$ nanostructures at 8 keV incident x-ray energy. The inset of fig 2(b) shows GIXRD profile of the nanostructure measured using a BRUKER system at Cu-K$\alpha$ radiation. (c) Measured TEM image for the nano-particles deposited on Cu grid. (d) Computed x-ray field intensity contour plot at 8 keV incident x-ray energy on top of a Si substrate.

## CONCLUSIONS

We have synthesized TiO$_2$ nanostructures using gas phase laser pyrolysis process. Nearly mono-dispersed, high pure TiO$_2$ nanoparticles of ~ 7-10 nm sizes were obtained on large area Si substrate. Depth-resolved x-ray fluorescene studies of TiO$_2$ NP have been performed by combined XRR and GIXRF techniques. The XSW analysis revealed that the TiO$_2$ nanoparticles exist in the form of an embedded layer on a thin TiO$_2$ layer medium on top of the substrate. The size distribution of the TiO$_2$ nanoparticles has also been confirmed by the TEM and GIXRD measurements. The methodology presented here has immense potential to determine reliable depth-resolved morphology and element-specific information about low dimensional nanomaterials.